# Direct Measurement of Magnetocaloric Effect in Metamagnetic $Ni_{43}Mn_{37.9}In_{12.1}Co_7$ Heusler Alloy

**A. P. Kamantsev[a, b], V. V. Koledov[a, b], A. V. Mashirov[a, b], E. T. Dilmieva[a], V. G. Shavrov[a], J. Cwik[b], and I. S. Tereshina[b, c]**

[a] *Kotelnikov Institute of Radio-Engineering and Electronics, Russian Academy of Sciences, Moscow, 125009 Russia*
[b] *International Laboratory of High Magnetic Fields and Low Temperatures, Wrocław, 53-421 Poland*
[c] *Baikov Institute of Metallurgy and Materials Science, Russian Academy of Sciences, Moscow, 119991 Russia*
*e-mail: kama@cplire.ru*

**Abstract**—The magnetocaloric effect in the metamagnetic $Ni_{43}Mn_{37.9}In_{12.1}Co_7$ Heusler alloy is directly studied experimentally under the adiabatic and quasi-isothermal conditions in a magnetic field with induction of up to 14 T.

## INTRODUCTION

In recent years, Heusler alloys, which exhibit a wide variety of interacting structural and magnetic phase transitions (PTs), have attracted much attention from researchers [1–3]. Of specific interest are studies of the magnetocaloric effect (MCE) in these alloys in connection with search of new methods for creation of solid-state magnetic refrigerators that operate at temperatures close to room temperature [4]. The MCE is generally treated as an adiabatic temperature change of a magnetic material when a magnetic field is applied to it (the $\Delta T$ effect). To develop a refrigerator based on such materials, however, it is important that we know the amount of heat that can be transferred from a sample with the MCE to a thermostat, or can be received from the thermostat under quasi-isothermal conditions in one cycle of heat transfer when the magnetic field is turned on or off (the $\Delta Q$ effect). Below, these quantities are referred to as the MCE under the adiabatic and quasi-isothermal conditions, respectively. Materials in which first-order structural PTs are accompanied by sharp changes in magnetic characteristics, are believed to hold the most promise. In these materials, high sensitivity of the temperature of the first-order structural PT to a magnetic field is observed, and is theoretically described by the Clausius–Clapeyron relation [3]. A first-order structural PT can coincide with both the Curie point (it is then called a magnetostructural PT) and the transition from antiferromagnetic to ferromagnetic ordering (it is then called a metamagnetostructural PT). As a rule, the MCEs in these materials have high absolute values and different signs [5, 6]; we then refer to them as "giant" direct and reverse MCEs. Among Heusler alloys, the metamagnetostructural PTs in Ni–Mn–In–Co alloys have record sensitivities to an external field (~10 K/T). In the alloys of this group, the reverse MCE is observed at temperatures close to that of the martensitic PT, since the martensitic phase is weakly magnetic (presumably antiferromagnetic), while the austenitic phase is ferromagnetic [6]. The current maximum reverse MCE measured directly in $Ni_{45.2}Mn_{36.7}In_{13}Co_{5.1}$ Heusler alloy is $\Delta T = -6.2$ K in a field with an induction of 2 T [7]. However, we have yet to find evidence that this $\Delta T$ is the maximum attainable absolute value for this alloy, since a magnetic field with an induction of 2 T is apparently too weak for a complete metamagnetostructural PT to occur over the field. In addition, there are not data on direct measurements of $\Delta Q$ in the literature practically. The aim of this work was to perfect the technique for measuring the $\Delta T$ and $\Delta Q$ effects in $Ni_{43}Mn_{37.9}In_{12.1}Co_7$ Heusler alloy with reverse MCEs in the vicinity of the metamagnetostructural transition in high magnetic fields (with induction of up to 14 T) and to make direct experimental measurements of these effects.

## EXPERIMENTAL

Samples of $Ni_{43}Mn_{50-x}In_xCo_7$ alloys in which $x =$ 12.1–12.5 at %, were produced via arc melting in argon atmosphere, followed by 48 h of homogenizing annealing at a temperature of 1173 K. Our object of study was the metamagnetic $Ni_{43}Mn_{37.9}In_{12.1}Co_7$ alloy, in which the temperatures of the start and finish of the direct and reverse martensitic transitions were (according to our data) $M_s = 285$ K and $M_f = 275$ K; $A_s = 304$ K and $A_f = 321$ K, respectively. The Curie point was $T_C = 430$ K. This alloy was selected for its temperature $M_f > 273$ K, which was convenient for conducting experiments with a water–ice thermostat.

To measure the $\Delta T$ effect, a vacuum calorimeter was developed that was placed in the field induced by a Bitter magnet. Samples with masses of 1–5 g were placed in a vacuum chamber, which was then evacuated to a pressure of ~$4 \times 10^{-4}$ kPa. The temperature was measured using diode sensors. We were able to measure $\Delta Q$ directly because sample of the material with the MCE was in good thermal contact with a massive block made of a nonmagnetic material that had a known specific heat and good thermal conductivity (Fig. 1). The amount of heat $\Delta Q$ transferred from the magnetic sample to the nonmagnetic block under quasi-isothermal conditions was determined by measuring the temperature change of the block due to magnetic field change. To ensure quasi-isothermal conditions of heat transfer and to neglect the amount of heat due to changes in the temperature of a sample, its mass $m$ is selected to be 10–20 times lower than that of the block $M$ ($m \ll M$). If we assume the external conditions to be adiabatic, neglect the heating of the sample and block by eddy currents and the work of magnetic field $H$ on the reverse magnetization of the sample in the closed cycle of turning the field on and off, we can write

$$\Delta Q + \Delta U = 0, \quad (1)$$

where $\Delta U = MC_b\Delta T_b$ is the change in the internal energy of the nonmagnetic block and $C_b$ and $\Delta T_b$ are the specific heat of the block and the change of its temperature due to magnetic field change, respectively. The value of the specific MCE $\Delta q$ under quasi-isothermal conditions can thus be expressed by the following approximate relation:

$$|\Delta q| = |\Delta Q|m^{-1} \approx Mm^{-1}C_b\Delta T_b. \quad (2)$$

This relation is valid with an error of 10–20% (the error grows as the induction of the field grows). In fields with inductions of >3 T, the specific work of the magnetic field does not exceed 150 J/kg (according to our data on magnetization). Heating by eddy currents is proportional to the squared rate of the increase in the magnetic field (8 or 14 T/min) and, with allowance for the shape of the copper block, can be estimated at no higher than 50 J/kg, while the intrinsic heating of the sample introduces no appreciable error because of its low mass, relative to that of the copper block.

During an experiment to measure the $\Delta T$ of a specimen, the vacuum calorimeter was put into a water–ice thermostat placed in the field of a Bitter magnet with an induction of up to 8 T. The initial temperature of the sample was set and then the measurements of this temperature began with simultaneous change of the magnetic field at a rate of 8 T/min. Figure 2 shows time dependences $\Delta T(t)$ and $H(t)$. The maximum absolute value of the sample's temperature change ($\Delta T = -3.3$ K) was recorded in the magnetic field at an induction of 8 T at an initial temperature of 273 K.

To measure $\Delta Q$, a sample with mass $m = 0.340$ g was glued to a massive copper block with mass $M$ = 4.523 g. Figure 3 shows the typical experimental time dependence for the change in the temperature of the copper block. Figure 4 shows dependence $\Delta q(H)$ obtained over several cycles of turning the magnetic field on and off at an initial temperature of the specimen of 273 K. It can be seen from this dependence that raising the magnetic field induction from 0 to 4 T has no noticeable effect. With a further increase in the magnetic field induction, the temperature of the block $T_b$ begins to fall; i.e., the reverse MCE occurs. The physical significance of this effect is that the austenite–martensite PT corresponds to an increase in the internal energy of the magnetic sample—which, under adiabatic conditions (when no heat from the environment is added to the PT's latent heat), is possible only if the temperature of the new phase falls. Applying the magnetic field leads to occurrence of a ferromagnetic austenitic phase (with a lower temperature) from the antiferromagnetic martensitic phase. The complete transformation of martensite into austenite occurs at a magnetic field induction of $\mu_0 H = 8$ T. Raising $\mu_0 H$ from 8 to 14 T increases the temperatures of the sample and block, but lowers the absolute value of $\Delta q$ from 2000 to 1600 J/kg. The reason for the change in the sign of the effect at $\mu_0 H > 8$ T could be the direct MCE in the ferromagnetic austenitic phase of the alloy. The backward branch of the magnetic cycle (from 14 to 4 T) is accompanied by a drop in the temperature of the sample, which then changes into a temperature increase. This is a metamagnetostructural transition in which a martensitic phase with a higher temperature is formed due to the reduction of the field induction under adiabatic conditions.

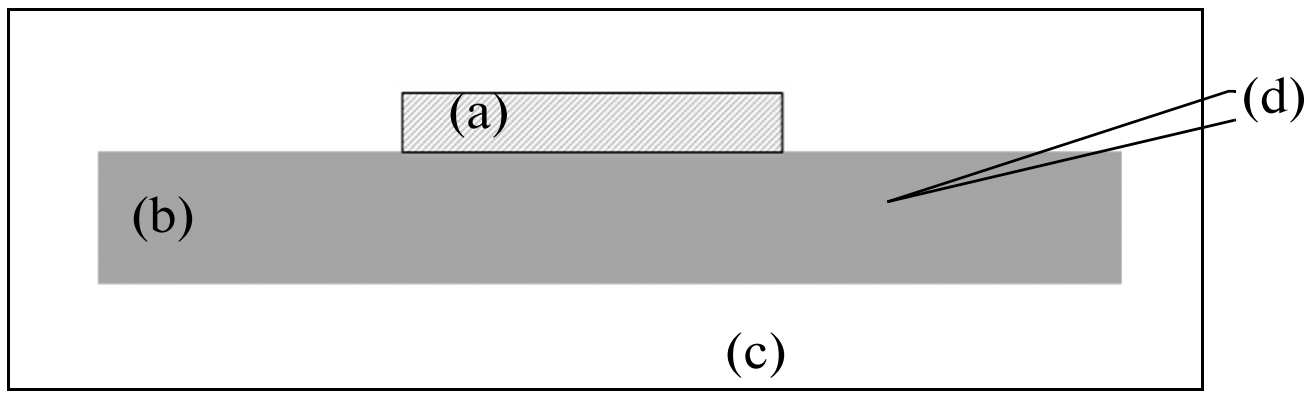

**Fig. 1.** Scheme of the direct experimental measurement of the MCE under quasi-isothermal conditions (the $\Delta Q$ effect): (a) sample with the MCE; (b) massive copper block; (c) vacuum chamber; and (d) diode temperature sensor.

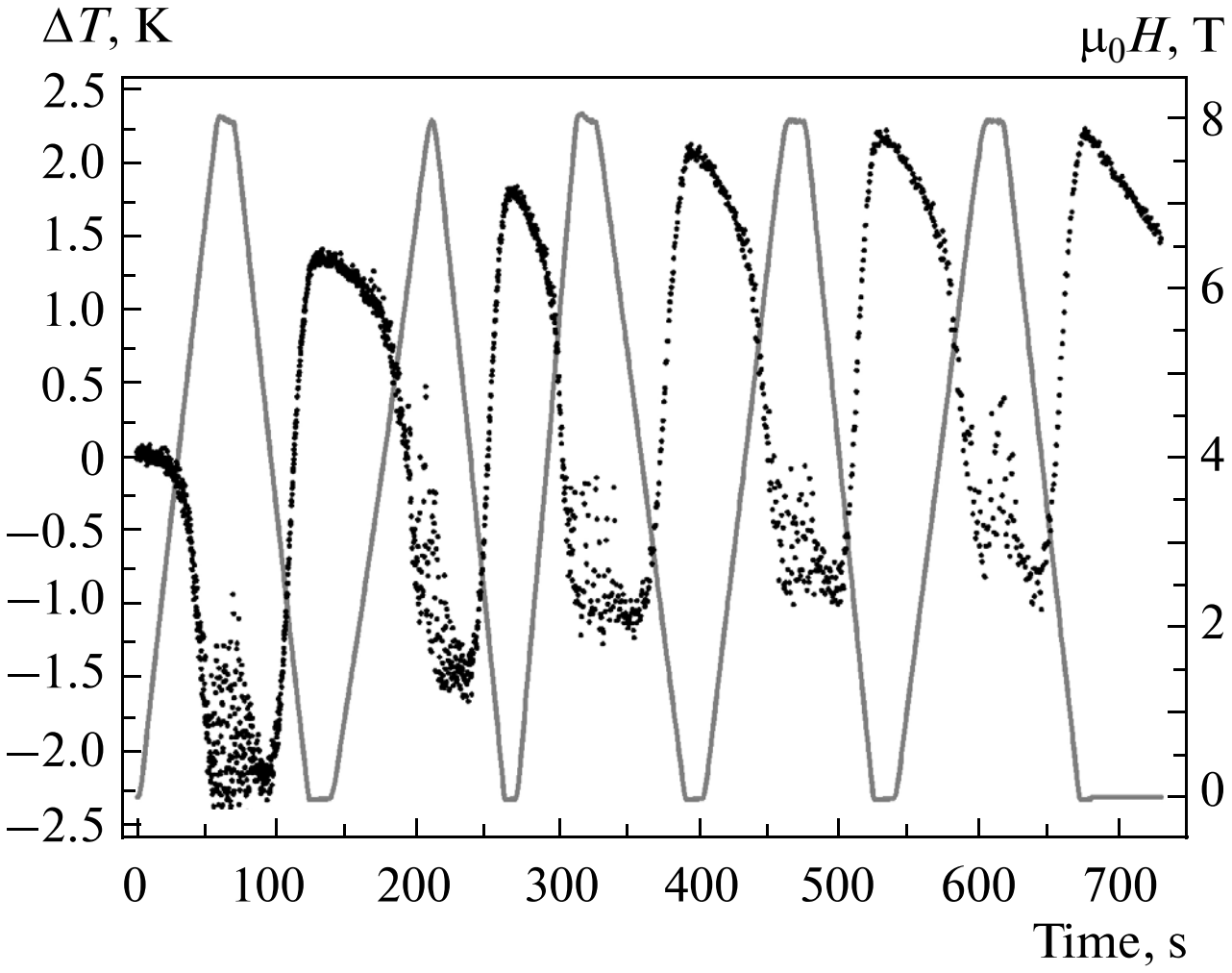

**Fig. 2.** Time dependences of the change in the magnetic field induction and the temperature change of the sample under adiabatic conditions (the $\Delta T$ effect). The initial temperature of the sample was 273 K.

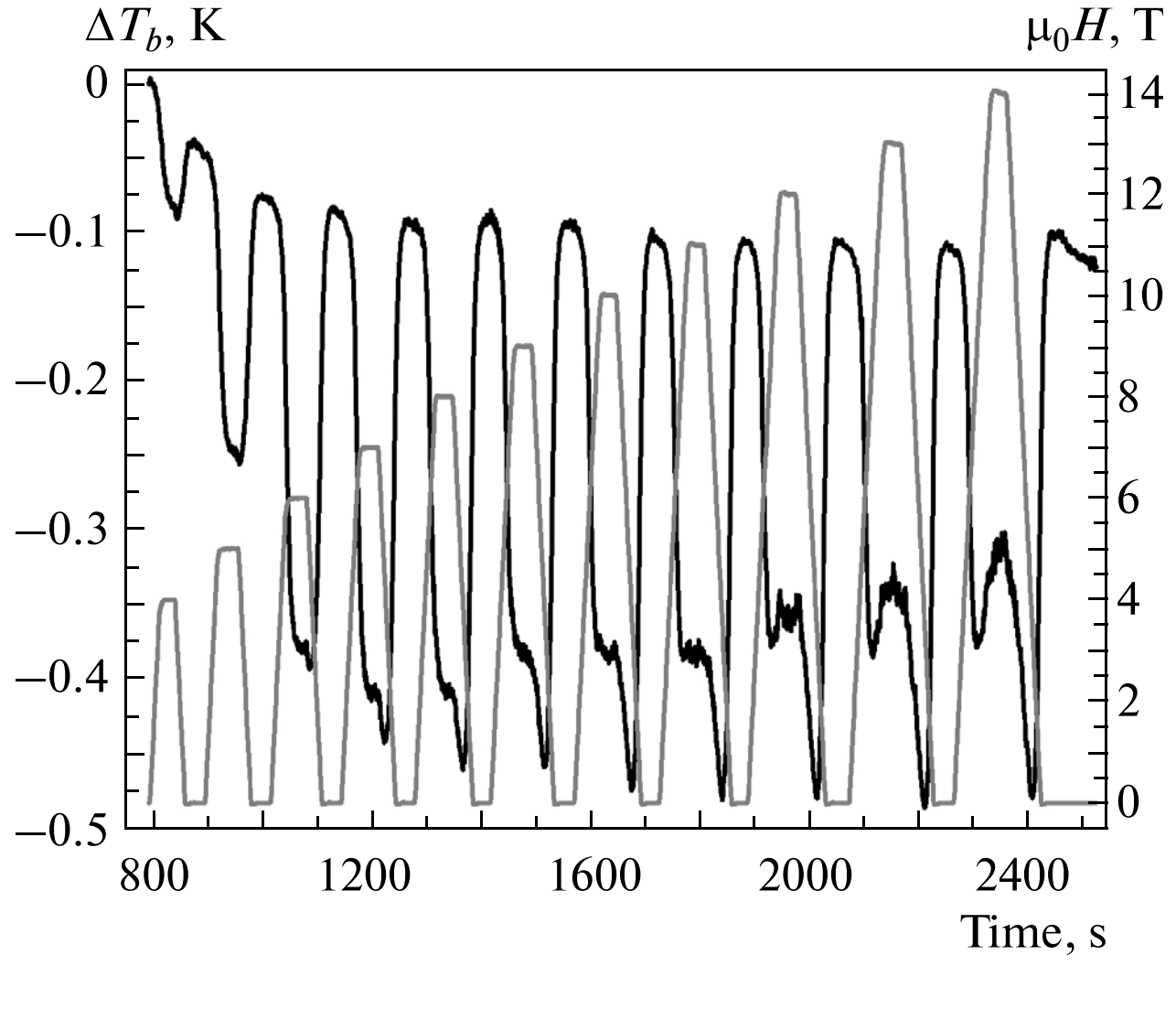

**Fig. 3.** Time dependences of the change in the magnetic field induction and the change in the temperature of the sample under quasi-isothermal conditions (the $\Delta Q$ effect). The initial temperature of the block and sample was 273 K.

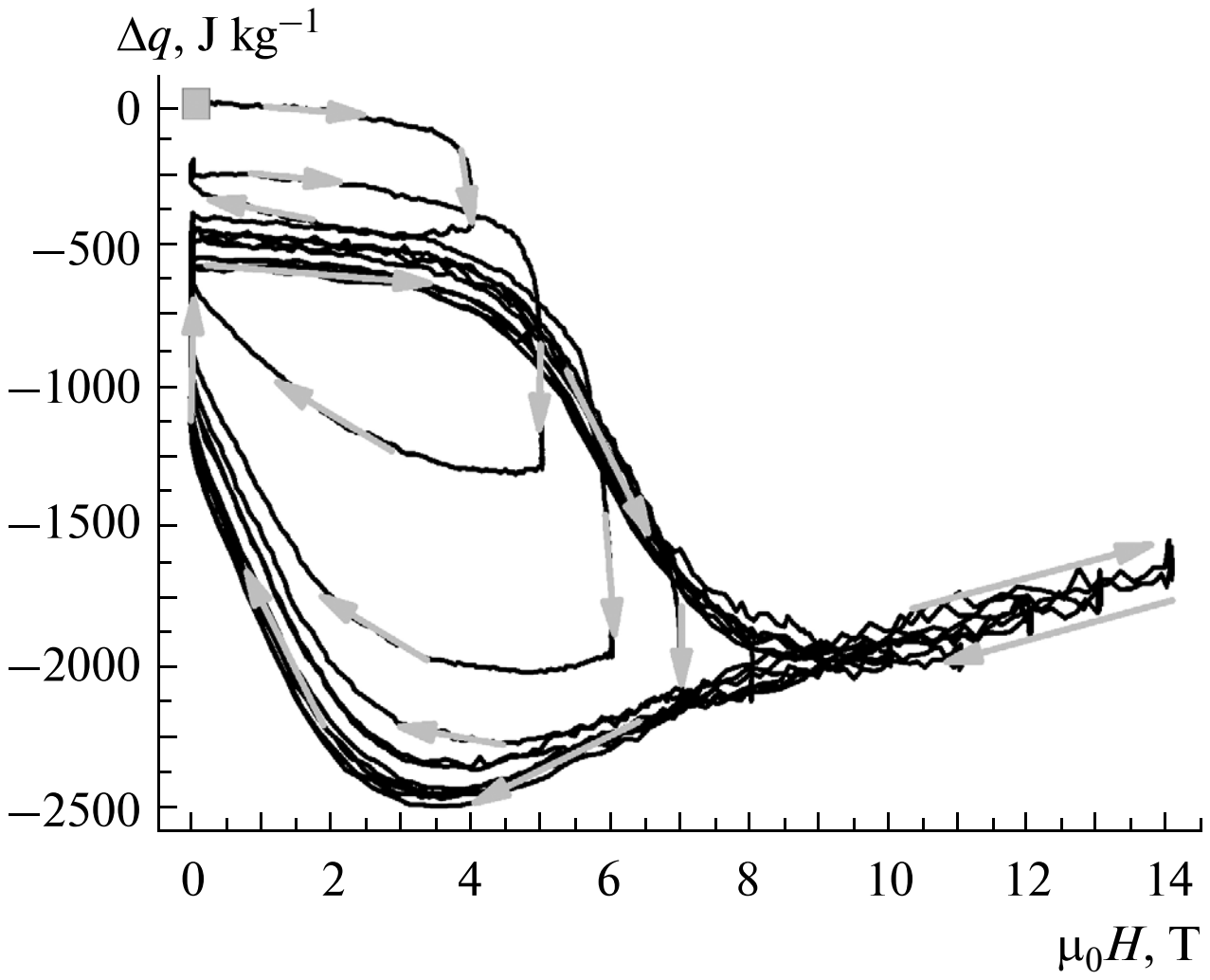

**Fig. 4.** Dependence of the MCE $\Delta q(H)$ under quasi-isothermal conditions on the magnetic field induction. Arrows show the direction of magnetic field change.

## CONCLUSIONS

Direct measurements of the MCE in the metamagnetic $Ni_{43}Mn_{37.9}In_{12.1}Co_7$ Heusler alloy under the adiabatic and quasi-isothermal conditions were made using an original technique. The reverse MCE reached its maximum values ($\Delta T = -3.3$ K and $\Delta q = -2000$ J/kg) at $\mu_0 H = 8$ T and an initial temperature of the samples of 273 K. Raising the magnetic induction field from 8 to 14 T reduced the absolute value of the MCE.

## ACKNOWLEDGMENTS

This work was supported by the Russian Foundation for Basic Research, project Nos. 12-08-01043, 12-07-00656, and 13-07-12130.